\documentclass[aps,prd,preprint,superscriptaddress,showpacs]{revtex4}
\usepackage{graphicx,dcolumn,url}
\def \Et {{E}_{T}}
\def \met  {\,/\!\!\!\!E_{T}}
\def\metsig{\not\!\!{E}_{T}^{sig}}
\def \ztt {Z^0 \rightarrow \tau^+\tau^-}
\def \ww  {W^+W^-}
\def \tt  {t\bar{t}}
\begin{document} 
%\pagewiselinenumbers
                                            
\title{
% \begin{flushright}
% {\small /CDF/DOC/TOP/PUBLIC/8300 \\ PRD-RC v2 \\ \today\\ }
% \end{flushright}
Cross Section Measurements of High-$p_T$ Dilepton Final-State Processes Using 
a Global Fitting Method }
\affiliation{Institute of Physics, Academia Sinica, Taipei, Taiwan 11529, Republic of China} 
\affiliation{Argonne National Laboratory, Argonne, Illinois 60439} 
\affiliation{Institut de Fisica d'Altes Energies, Universitat Autonoma de Barcelona, E-08193, Bellaterra (Barcelona), Spain} 
\affiliation{Baylor University, Waco, Texas  76798} 
\affiliation{Istituto Nazionale di Fisica Nucleare, University of Bologna, I-40127 Bologna, Italy} 
\affiliation{Brandeis University, Waltham, Massachusetts 02254} 
\affiliation{University of California, Davis, Davis, California  95616} 
\affiliation{University of California, Los Angeles, Los Angeles, California  90024} 
\affiliation{University of California, San Diego, La Jolla, California  92093} 
\affiliation{University of California, Santa Barbara, Santa Barbara, California 93106} 
\affiliation{Instituto de Fisica de Cantabria, CSIC-University of Cantabria, 39005 Santander, Spain} 
\affiliation{Carnegie Mellon University, Pittsburgh, PA  15213} 
\affiliation{Enrico Fermi Institute, University of Chicago, Chicago, Illinois 60637} 
\affiliation{Comenius University, 842 48 Bratislava, Slovakia; Institute of Experimental Physics, 040 01 Kosice, Slovakia} 
\affiliation{Joint Institute for Nuclear Research, RU-141980 Dubna, Russia} 
\affiliation{Duke University, Durham, North Carolina  27708} 
\affiliation{Fermi National Accelerator Laboratory, Batavia, Illinois 60510} 
\affiliation{University of Florida, Gainesville, Florida  32611} 
\affiliation{Laboratori Nazionali di Frascati, Istituto Nazionale di Fisica Nucleare, I-00044 Frascati, Italy} 
\affiliation{University of Geneva, CH-1211 Geneva 4, Switzerland} 
\affiliation{Glasgow University, Glasgow G12 8QQ, United Kingdom} 
\affiliation{Harvard University, Cambridge, Massachusetts 02138} 
\affiliation{Division of High Energy Physics, Department of Physics, University of Helsinki and Helsinki Institute of Physics, FIN-00014, Helsinki, Finland} 
\affiliation{University of Illinois, Urbana, Illinois 61801} 
\affiliation{The Johns Hopkins University, Baltimore, Maryland 21218} 
\affiliation{Institut f\"{u}r Experimentelle Kernphysik, Universit\"{a}t Karlsruhe, 76128 Karlsruhe, Germany} 
\affiliation{High Energy Accelerator Research Organization (KEK), Tsukuba, Ibaraki 305, Japan} 
\affiliation{Center for High Energy Physics: Kyungpook National University, Taegu 702-701, Korea; Seoul National University, Seoul 151-742, Korea; and SungKyunKwan University, Suwon 440-746, Korea} 
\affiliation{Ernest Orlando Lawrence Berkeley National Laboratory, Berkeley, California 94720} 
\affiliation{University of Liverpool, Liverpool L69 7ZE, United Kingdom} 
\affiliation{University College London, London WC1E 6BT, United Kingdom} 
\affiliation{Centro de Investigaciones Energeticas Medioambientales y Tecnologicas, E-28040 Madrid, Spain} 
\affiliation{Massachusetts Institute of Technology, Cambridge, Massachusetts  02139} 
\affiliation{Institute of Particle Physics: McGill University, Montr\'{e}al, Canada H3A~2T8; and University of Toronto, Toronto, Canada M5S~1A7} 
\affiliation{University of Michigan, Ann Arbor, Michigan 48109} 
\affiliation{Michigan State University, East Lansing, Michigan  48824} 
\affiliation{Institution for Theoretical and Experimental Physics, ITEP, Moscow 117259, Russia} 
\affiliation{University of New Mexico, Albuquerque, New Mexico 87131} 
\affiliation{Northwestern University, Evanston, Illinois  60208} 
\affiliation{The Ohio State University, Columbus, Ohio  43210} 
\affiliation{Okayama University, Okayama 700-8530, Japan} 
\affiliation{Osaka City University, Osaka 588, Japan} 
\affiliation{University of Oxford, Oxford OX1 3RH, United Kingdom} 
\affiliation{University of Padova, Istituto Nazionale di Fisica Nucleare, Sezione di Padova-Trento, I-35131 Padova, Italy} 
\affiliation{LPNHE, Universite Pierre et Marie Curie/IN2P3-CNRS, UMR7585, Paris, F-75252 France} 
\affiliation{University of Pennsylvania, Philadelphia, Pennsylvania 19104} 
\affiliation{Istituto Nazionale di Fisica Nucleare Pisa, Universities of Pisa, Siena and Scuola Normale Superiore, I-56127 Pisa, Italy} 
\affiliation{University of Pittsburgh, Pittsburgh, Pennsylvania 15260} 
\affiliation{Purdue University, West Lafayette, Indiana 47907} 
\affiliation{University of Rochester, Rochester, New York 14627} 
\affiliation{The Rockefeller University, New York, New York 10021} 
\affiliation{Istituto Nazionale di Fisica Nucleare, Sezione di Roma 1, University of Rome ``La Sapienza," I-00185 Roma, Italy} 
\affiliation{Rutgers University, Piscataway, New Jersey 08855} 
\affiliation{Texas A\&M University, College Station, Texas 77843} 
\affiliation{Istituto Nazionale di Fisica Nucleare, University of Trieste/\ Udine, Italy} 
\affiliation{University of Tsukuba, Tsukuba, Ibaraki 305, Japan} 
\affiliation{Tufts University, Medford, Massachusetts 02155} 
\affiliation{Waseda University, Tokyo 169, Japan} 
\affiliation{Wayne State University, Detroit, Michigan  48201} 
\affiliation{University of Wisconsin, Madison, Wisconsin 53706} 
\affiliation{Yale University, New Haven, Connecticut 06520} 
\author{A.~Abulencia}
\affiliation{University of Illinois, Urbana, Illinois 61801}
\author{J.~Adelman}
\affiliation{Enrico Fermi Institute, University of Chicago, Chicago, Illinois 60637}
\author{T.~Affolder}
\affiliation{University of California, Santa Barbara, Santa Barbara, California 93106}
\author{T.~Akimoto}
\affiliation{University of Tsukuba, Tsukuba, Ibaraki 305, Japan}
\author{M.G.~Albrow}
\affiliation{Fermi National Accelerator Laboratory, Batavia, Illinois 60510}
\author{D.~Ambrose}
\affiliation{Fermi National Accelerator Laboratory, Batavia, Illinois 60510}
\author{S.~Amerio}
\affiliation{University of Padova, Istituto Nazionale di Fisica Nucleare, Sezione di Padova-Trento, I-35131 Padova, Italy}
\author{D.~Amidei}
\affiliation{University of Michigan, Ann Arbor, Michigan 48109}
\author{A.~Anastassov}
\affiliation{Rutgers University, Piscataway, New Jersey 08855}
\author{K.~Anikeev}
\affiliation{Fermi National Accelerator Laboratory, Batavia, Illinois 60510}
% ...
\collaboration{CDF Collaboration\footnote{With visitors from $^a$University of Athens, 
$^b$University of Bristol, 
$^c$University Libre de Bruxelles, 
$^d$Cornell University, 
$^e$University of Cyprus, 
$^f$University of Dublin, 
$^g$University of Edinburgh, 
$^h$University of Heidelberg, 
$^i$Universidad Iberoamericana, 
$^j$University of Manchester, 
$^k$Nagasaki Institute of Applied Science, 
$^l$University de Oviedo, 
$^m$University of London, Queen Mary and Westfield College, 
$^n$Texas Tech University, 
$^o$IFIC(CSIC-Universitat de Valencia), 
}}
\noaffiliation

\date{\today}

\begin{abstract}
We present a new method for studying high-$p_T$ dilepton events ($e^{\pm}e^{\mp}$,
$\mu^{\pm}\mu^{\mp}$, $e^{\pm}\mu^{\mp}$) and simultaneously extracting the 
production cross sections of $p\bar{p} \to t\bar{t}$, $p\bar{p} \to W^+W^-$, and 
$p\bar{p} \to \ztt$ at a center-of-mass energy of $\sqrt{s} = 1.96\,{\rm TeV}$. 
We perform
a likelihood fit to the dilepton data in a parameter space defined by the
missing transverse energy and the number of jets in the event. Our
results, which use $360\,{\rm pb^{-1}}$ of data recorded with the CDF II detector at 
the Fermilab Tevatron Collider,
are $\sigma(t\bar{t}) = 8.5_{-2.2}^{+2.7}\,\rm{pb}$,
$\sigma(W^+W^-) = 16.3^{+5.2}_{-4.4}\,\rm{pb}$, and $\sigma(\ztt)
=291^{+50}_{-46}\,\rm{pb}$. 
%This method utilizes the full statistical
%power of the dilepton sample for given lepton definitions, and provides a 
%global test of standard
%model predictions for the high-$p_{T}$ dilepton final state.
\end{abstract}
                           

\maketitle

%--------------
% introduction
% ------------

%High-$p_{T}$ dilepton signatures provide a distinct way to test the

%standard model and physics beyond. There are relatively few standard
%model processes with a final state containing a highly energetic
%electron and muon, and for the main such processes, other
%characteristics of the events are very different.

There are relatively few standard model (SM) processes that 
contribute significantly to a final-state containing a pair of highly
energetic charged leptons. The processes that can contribute to
these ``dilepton'' events include top-quark pair ($\tt$)
production, $W$-boson pair ($\ww$) production, and Drell-Yan processes. 
The distinctiveness of this final-state offers unique tests of the SM and 
an intriguing potential for revealing new physics. For instance, in Run I 
(1992-1996) the Collider Detector at Fermilab (CDF) observed several 
$\tt$ candidate events in the dilepton decay mode~\cite{dilcross}
with unusual characteristics, and it was suggested that
the kinematics of these events could be better described by the
cascade decays of heavy supersymmetric quarks~\cite{barney}.
%Standard model (SM) processes with a final-state containing a pair of highly
%energetic charged leptons provide a distinct way to test the SM and
%search for new physics. 
%In Run I (1992-1996) the Collider Detector at Fermilab (CDF) observed several 
%top quark pair ($\tt$) candidate events in the dilepton decay mode~\cite{dilcross}
%with large $\met$, and large lepton $p_T$, and it was suggested that
%the kinematics of these events could be better described by the
%cascade decays of heavy supersymmetric quarks~\cite{barney}. 
Furthermore, the top-quark's extraordinarily large mass might be an indication
of a close connection with the mechanism of mass generation itself, as
for example, in the model of topcolor assisted technicolor~\cite{topcolor}, which
predicts new resonances decaying to $\tt$.  
In addition, a fourth generation fermion family~\cite{4gen} could enhance
the gluon-gluon fusion Higgs production cross section by an order of
magnitude, which, for a heavy Higgs boson, would lead to an increase
in the number of $W$ boson pairs observed~\cite{hww}. These examples
could all produce an excess of dilepton events in our data, which, depending on
the topology of the new physics, would affect the
cross section measurements of the main SM processes to different extents.
This provides the motivation behind the present study of highly energetic
dilepton events.

The main SM processes with a high-$p_T$~\cite{peetee} dilepton final-state can
be identified based on their distinct event characteristics. 
For example, $\tt$ and $\ww$ production with decays 
in the $e\mu$ dilepton channel, 
$t\bar{t} \rightarrow W^+b\,W^-\bar{b} \rightarrow e^{\pm}\mu^{\mp}\,
\nu\bar{\nu}\,b\bar{b}$ and $ W^+W^- \rightarrow e^{\pm}\mu^{\mp}\,
\nu\bar{\nu}$, and the di-tau decays of $Z^0$ bosons, $Z^0 \rightarrow
\tau^+ \tau^- \rightarrow e^{\pm}\mu^{\mp} \nu\nu\bar{\nu}\bar{\nu}$, 
can be distinguished from each other by the number of
jets, $N_{j}$, and the missing transverse energy~\cite{met},
$\met$, in the event.
%can all
%produce final states with two high-$p_{T}$~\cite{peetee} leptons. However, these
%processes are
%very distinct from each other when one considers the number of
%jets~\cite{jetdef}, $N_{j}$, and missing transverse energy~\cite{met},
%$\met$, expected in the event. 
Both $t\bar{t}$ and $W^+W^-$ events
typically have large $\met$ from the final-state undetected neutrinos; however,
due to the two final state $b$-quarks, $t\bar{t}$ has a greater number of jets
than $\ww$.  Conversely, $\ztt$ events have small $\met$ (due
to the neutrinos being of lower energy than in the $\tt$ and $\ww$ processes, and
typically traveling in opposite directions), and most often no jets. For both
$\ww$ and $\ztt$ events, jets can arise only through higher-order
processes that include initial-state gluon radiation.

In contrast to analyses dedicated to measuring a single SM process, 
the analysis presented here adopts a more global strategy by 
considering all events with a high-$p_{T}$ electron and muon, and making no 
further selection requirements.
We then exploit the different $\met$ and
$N_j$ characteristics to simultaneously extract the production
cross sections of the three main processes described. This is done by
fitting the $e\mu$ data in a two-dimensional (2-D) $\met - N_j$ parameter space to 
template distributions of
$\tt$, $\ww$, and $\ztt$ events. Less significant processes
are also taken into account.  

We also consider the $ee$ and $\mu\mu$ final-states, in addition to $e\mu$,
but in these cases we have the added complication of a large 
Drell-Yan ($Z^0/\gamma^\ast \rightarrow
e^+e^- {\rm \ or\ } \mu^+\mu^-$) contribution, necessitating a
different treatment for these channels. This involves
reducing the Drell-Yan $ee$ and $\mu\mu$ contributions by requiring 
events to have significant $\met$ in those channels. Without 
this requirement the $\tt$ and $\ww$ contributions to the $ee$ and $\mu\mu$ events
would be overwhelmed, rendering these final-states unusable.
Our $\tt$ and $\ww$ results use all three dilepton final-states. 
For extracting the $\ztt$ cross section we use only $e\mu$ events
as the removal of Drell-Yan $ee$ and $\mu\mu$ events
also significantly reduces the $\ztt$ contribution in the $ee$ and $\mu\mu$ channels.

Since this method makes minimal requirements on events after
requiring two leptons, it utilizes the full statistical power of the data for
given lepton definitions.
% thereby
%providing the potential for simultaneously measuring the $\tt$, $\ww$
%and $\ztt$ cross sections with the greatest precision for the lepton
%definitions used. 
In the present
analysis we have chosen to use very tight lepton identification requirements
%which afford us greater control of the background processes, however
%which somewhat reduce the potential gain from this method. 
%We do so in order
to demonstrate the method, and establish a foundation for future measurements
with more data.
This gives us greater control of the background processes, in particular those
involving jets being misidentified as, or containing, leptons. However,
it also reduces the potential statistical gain from the method. 
Even so, the results are comparable in precision to the analyses 
dedicated to the individual cross section measurements, and that use looser
dilepton definitions. 
Furthermore, by looking at all processes simultaneously
in the same generic dilepton sample, this method tests the SM consistency in
a way an analysis dedicated to a single cross section measurement does not.
An additional benefit
of minimizing the requirements on the event after two high-$p_T$
leptons are selected is the possibility for sensitivity to new physics
which might fall into the $\met - N_j$ parameter space that we study.

The data sample was collected with the CDF\,II detector between 2002
and 2004, and corresponds to an integrated luminosity of approximately
360 ${\rm pb^{-1}}$. The results presented here complement the
individual cross section measurements from CDF for each
process~\cite{ttdilxsec, WWxsec, zttxsec}, which make additional requirements
to reduce, and assume the SM cross sections for all the processes in the dilepton 
sample other than the one being measured.

%---------
% detector
% --------
The CDF II detector has a general-purpose design~\cite{cdfdet}. The components 
relevant to this analysis are briefly 
described here. A tracking system inside a 1.4~T superconducting solenoidal magnet
is composed of silicon detectors for high-precision track measurements,
and an open-cell drift chamber surrounding the silicon system.  Electromagnetic and
hadronic calorimeters surround the tracking system and solenoid and
are used to measure the energy of interacting particles. In this analysis information
is used from both the central (covering the pseudo-rapidity range $|\eta|<1.1$)
and end-plug detectors (of which we use the range $1.2 <|\eta|< 2.0$) 
to identify electron
candidates. The missing transverse energy calculation uses calorimeter towers with
$|\eta|<3.5$. We identify jets using clusters of calorimeter towers above an 
energy threshold of 3 GeV, with fixed cone radius 
$\Delta R=\sqrt{{\Delta\phi}^{2}+ {\Delta \eta}^{2}}= 0.4$. The jet 
transverse energy is corrected for the
calorimeter response and multiple interactions~\cite{jetcorr, jes}. We
require that the corrected jets have $E_{T}> 15~{\rm GeV}$ and $|\eta|<2.5$.
A set of
drift chambers located outside the central hadronic calorimeters are used to
detect muons in the region  $|\eta| < 0.6$. 
%Muons are minimum ionizing particles
%and therefore do not deposit a significant fraction of their energy in the calorimeters. 
Additional drift chambers and scintillation counters 
detect muons in the region $0.6 < |\eta| < 1.0$.

A three-level trigger system is used to select events.
The triggers employed to collect events for this analysis are~\cite{trigele}
an inclusive central electron
($|\eta|<1.1$) trigger requiring an electron with $E_{T} > 18~{\rm GeV}$,
and an inclusive central muon ($|\eta|<1.0$) trigger requiring a muon with
$p_{T}> 18~{\rm GeV}/c$.
% and a trigger for events with a forward
%electron ($1.2 < |\eta| < 2.8$) with 
%$E_{T}> 20~{\rm GeV}$ and $\met > 15~{\rm GeV}$.

%----------------
% event selection 
% ---------------

Events selected for the analysis contain two opposite sign
leptons (electrons or muons) consistent with originating from the same
vertex, and with $E_{T}> 20~{\rm GeV}$ for electrons
and $p_{T}> 20~{\rm GeV}/c$ for muons. Muons from
cosmic rays and electrons from photon conversions are removed, as described
in Ref.~\cite{trigele}. Both
leptons are required to be isolated in the
calorimeter~\cite{caliso} and the tracking chamber~\cite{trkiso}, in
order to reduce the probability that a jet is misidentified as a
lepton, or that a selected lepton comes from the semileptonic decay of 
$b$ or $c$ hadrons.

%For electrons with $|\eta| > 1.2$, this track-energy cluster association 
%utilizes a calorimeter seeded 
%silicon tracking algorithm~\cite{phoenix}.
%\bibitem{phoenix} C. Issever, AIP Conf. Proc. 670, 371 (2001). 

We consider three classes of dilepton events: $e\mu$, $ee$, and
$\mu\mu$.  For the $e\mu$ channel we make no further requirements on
the event after the selection of two leptons as described above, 
and simply count the number of jets, $N_j$, and measure the $\met$ in the event. 
%We then 
%study the $e\mu$ sample in a 2-D $\met - N_j$ parameter space.

The $ee$ and $\mu\mu$ channels have a large Drell-Yan contribution
%($Z^0/\gamma^\ast \rightarrow ee/\mu\mu$)
which we reduce significantly by applying a requirement on the missing transverse energy 
significance defined by:
\[ \metsig = \frac{\met}{\sqrt{\sum E_T}} \]
where $\sum E_T$ is the sum of transverse energies over all calorimeter towers, 
corrected to include the $p_T$ of the muons. We require
$\metsig > 2.5\,{\rm GeV^{1/2}}$ for all $ee$ and $\mu\mu$ events .

%%%%%%%%%%%%%%% definition of signal and background processes

\begin{figure*}[h]
\includegraphics[height=14.6cm,width=\columnwidth]{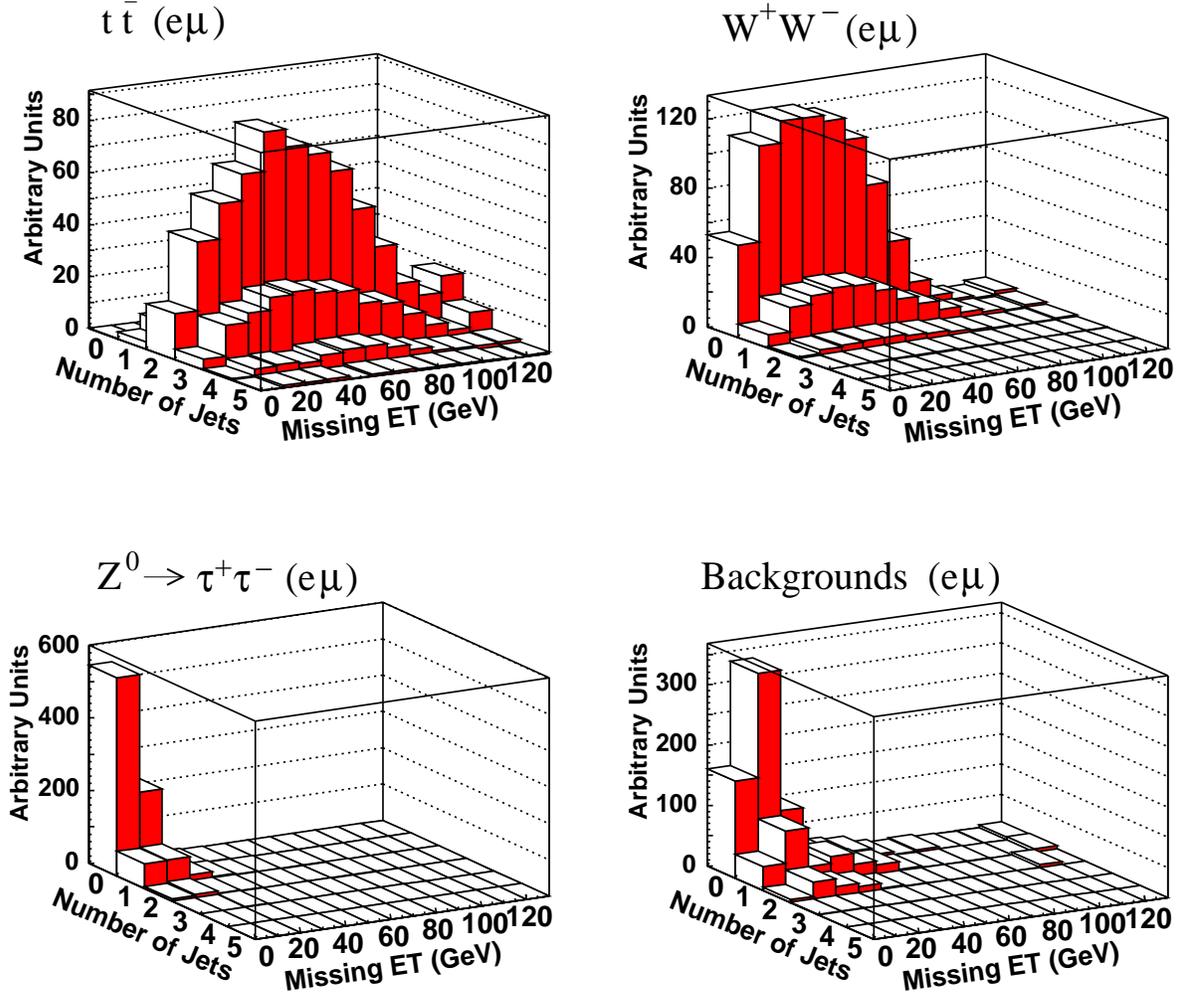}
\caption{\label{fig:fig1}
The $\met - N_{j}$ distributions for $\tt$, $W^{+}W^{-}$, $\ztt$ in the
$e\mu$ channel. Also shown is the combined background distribution
for the $e\mu$ channel. All the distributions 
are normalized to an arbitrary equal volume. The numbers of events expected 
in $360\,{\rm pb^{-1}}$ for each source are given in Table~\ref{table:gstable}.
The highest  $\met$ and  $N_{j}$ bins include any overflow events.}
\end{figure*}

After the above dilepton and $\metsig$ requirements, the dominant SM contributions 
which we consider as our signal processes, are $t\bar{t} \rightarrow W^+b\,W^-\bar{b}
\rightarrow \ell^+\nu b\, \ell^- \bar{\nu}\bar{b}$, 
$W^+W^- \rightarrow \ell^+\nu \, \ell^- \bar{\nu}$, and
$Z^0 \rightarrow \tau^+\tau^- \rightarrow \ell^+ \nu_{\ell} \bar{\nu}_{\tau}\,
\ell^- \bar{\nu}_{\ell} \nu_{\tau}$, where $\ell$ indicates an electron or muon.
These processes are separated in our chosen $\met - N_j$ parameter space
as previously explained.
%$t\bar{t}$ production with decay in the dilepton channel is typified by large
%$\met$ and large $N_j$ (mostly from the two final-state $b$-quarks); 
%$W^+W^-$ production with decay in the dilepton channel is characterized by
%large $\met$ and small $N_j$ (any jet activity is due to initial-state QCD
%radiation); and $Z^0 \rightarrow \tau^+\tau^-$ has typically small $\met$ and
%small  $N_j$. 
The predicted distributions of these signal processes in the 
$\met - N_j$ parameter space for the $e\mu$ channel are shown in Fig.~\ref{fig:fig1}, 
and were obtained from Monte Carlo simulations.
%These distributions were obtained 
%from Monte Carlo simulations, except for the $W+ {\rm jets}$ contribution to the 
%combined background distribution, which was extracted from data and is
%discussed further below.

%    figures to include \\
 %  -- templates for WW, tt, Z -> tau-tau, and data (maybe just for emu ??)\\}

\begin{figure*}[h]
\includegraphics[height=8.0cm]{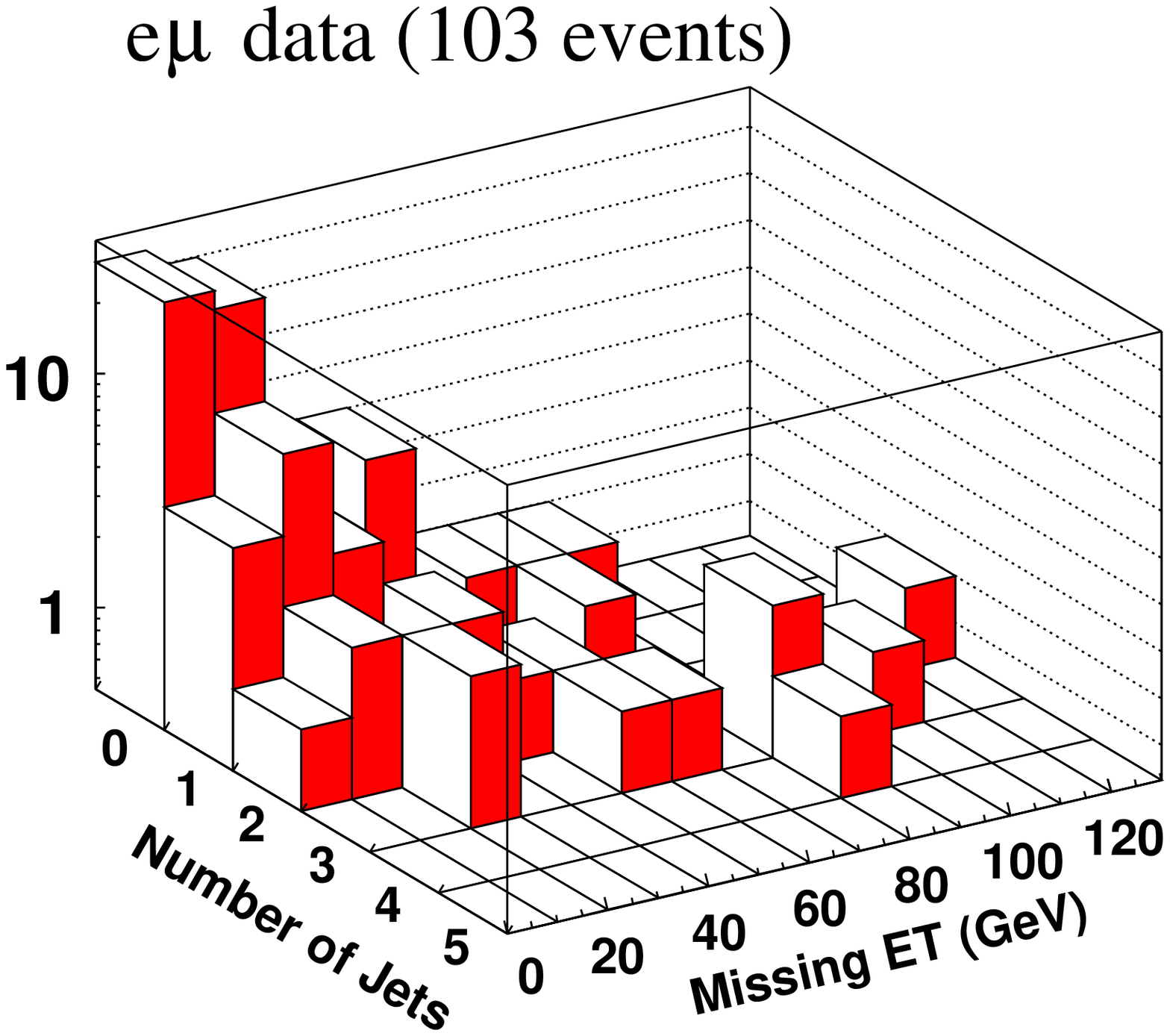}
\caption{\label{fig:fig2}
The $\met - N_{j}$ distribution for data in the $e\mu$ channel. 
The log scale is used to make the low count bins more visible.}
\end{figure*}

Contributions to the dilepton sample
%in this $\met - N_j$ parameter space, 
which we consider as background processes, include
Drell-Yan ($ee$, $\mu\mu$), $WZ$, $ZZ$, $W+\gamma$, and
$W+ {\rm jets}$. Note that Drell-Yan $\mu\mu$ events could be
reconstructed as an $e\mu$ final-state when a muon in the forward region
radiates a photon that is misidentified as an electron. 
The predicted combined background distributions in the 
$\met - N_j$ parameter space is also shown in Fig.~\ref{fig:fig1}.
We fit the data $e\mu$, $ee$, and $\mu\mu$
distributions to the expected signal shapes, letting each of their
normalizations float in the fit.
The $e\mu$ data distribution is shown in Fig.~\ref{fig:fig2}, of which
about 60\% of the events are expected to be from $\ztt$ and concentrated
in the low-$\met$, zero-jet region of the $\met - N_{j}$ parameter space.

%-------------------------
%efficiency and acceptance
%-------------------------
%{\bf efficiency and acceptance section here \\}
To determine the acceptance of our selection criteria for the $\tt$, $\ww$, and $\ztt$ 
processes we use the
{\sc pythia}~\cite{pythia} Monte Carlo program, followed by a full simulation of the
CDF II detector which is based on the {\sc geant} simulation program~\cite{geant}. 
The acceptances for each of these processes are
shown in Table~\ref{table:acctable}. The $Z^0 \rightarrow \tau^+\tau^-$ acceptances
are defined as the fraction of $Z^0/\gamma^\ast \rightarrow \tau^+\tau^-$ events
generated in the di-tau mass range of $66 < M_{\tau\tau} < 116\,{\rm GeV}/c^2$
that pass our dilepton selection criteria. The contribution from $\gamma^\ast$ is only
about 0.3\%.
For the $ee$ and $\mu\mu$ channels the
additional $\metsig$ requirement removes 32\% of $\tt$ events and 36\% of $\ww$
events, while also removing about 99.8\% of Drell-Yan events.

\begin{table}[h]
\caption{\label{table:acctable}
Summary of acceptances for $t\bar{t}$, $\ww$, and $\ztt$ events, where the quoted errors 
include the systematic uncertainties from Table~\ref{table:accsys}.
Values include SM branching
fractions to the dilepton final-state. 
The $t\bar{t}$ events were simulated with a top quark mass of 178 GeV/$c^{2}$.}
\begin{ruledtabular}
\begin{tabular}{lccc}
   &             $e\mu$ & $ee$ & $\mu\mu$  \\ \hline
$t\bar{t}$
    & $(0.399\pm 0.029)$\% & $(0.144\pm 0.019 )$\% & $(0.136\pm0.015)$\% 
      \\
 %   & $10.0\pm 0.7$ & $3.6\pm 0.5$ & $3.4\pm 0.4$ & $17\pm 1.6$  \\
$\ww$
    & $(0.294 \pm 0.018)$\% & $(0.111 \pm 0.008 )$\% & $(0.092 \pm 0.006 )$\% 
      \\
 %   & $13.8 \pm 0.8$ & $5.2\pm 0.4$ & $4.3 \pm 0.3$ & $23.3 \pm 1.5$  \\
$Z\rightarrow \tau \tau$
    & $(0.0458 \pm 0.0032 )$\% & $(0.0008 \pm 0.0001 )$\% & $(0.0005 \pm 0.0001)$\% 
       \\
 %   & $57.8\pm 4$ & $1.1\pm 0.2$ & $0.6\pm 0.1$ & $59.9\pm 4.3$  \\
\end{tabular}
\end{ruledtabular}
\end{table}

The systematic uncertainties on the $\tt$, $\ww$, and $\ztt$
acceptances, summarized in Table~\ref{table:accsys}, result from
uncertainties in the jet energy scale (JES)~\cite{jes}, the modeling of the
initial-state radiation (ISR) by {\sc pythia} [and for the case of
$t\bar{t}$ also final-state radiation (FSR)], the uncertainty on the
parton distribution functions~\cite{pdferror}, the modeling of $\metsig$, and uncertainties
in the lepton trigger and identification efficiencies. The $\metsig$
systematic uncertainty does not apply to the $\ztt$ cross section measurement, as we
fit for this only in the $e\mu$ channel. Moreover, the $e\mu$
channel acceptance does not suffer from a jet energy scale uncertainty, as we
include $e\mu$ events with all jet multiplicities, whereas for the
$ee$ and $\mu\mu$ channels this systematic uncertainty enters through the
$\metsig$ requirement.
%In addition there is an
%uncertainty on the integrated luminosity (6$\%$), which applies to
%signal and any backgrounds. 
A 6\% uncertainty on the integrated 
luminosity is applied to the expected number of events for all processes~\cite{lum}.

\begin{table}[h]
\caption{
Summary of systematic uncertainties on the acceptance for each ``signal''
process. See text for further details.
\label{table:accsys}
}
\begin{ruledtabular}
\begin{tabular}{lccccccc}
                                                                                
Source & $\tt$($ee$) & $\tt$($e\mu$) & $\tt$($\mu\mu$)
&  $\ww$($ee$) & $\ww$($e\mu$) & $\ww$($\mu\mu$) & $\ztt$
($e\mu$) \\
\hline
  JES & $5\%$ & - & $6\%$ & $1\%$ & - & $1\%$ & - \\
                                                                                
  ISR  & $8\%$ & $4\%$ & $6\%$ & $5\%$ & $5\%$ & $5\%$ & $5\%$ \\
                                                                                
  FSR & $7\%$ & $3\%$ & $5\%$ & - & - & - & - \\
                                                                                
  Other & $6\%$ & $5\%$ & $5\%$ & $5\%$ & $4\%$ & $4\%$ & $5\%$ \\
                                                                                
\hline
 Total  & $13\%$ & $7\%$ & $11\%$ & $7\%$ & $6\%$ & $7\%$ & $7\%$\\
                                                                                
\end{tabular}
\end{ruledtabular}
\end{table}
                                                                                
%------------
% backgrounds
%------------   
The Drell-Yan and diboson ($WZ, ZZ$)
backgrounds are determined using {\sc pythia} Monte Carlo, followed by the
detector simulation. We normalize the total number of events
for these processes to theoretical
cross section predictions~\cite{wwxsec}. To estimate the $W+\gamma$ background
we use a matrix element generator~\cite{baur1} and use {\sc pythia} for the
initial-state QCD radiation and hadronization.
The background from $W+ {\rm jets}$, where a jet or track is 
misidentified as an electron or muon, is determined from the data.
We first calculate the
probability that a jet with a large fraction of its energy deposited
in the electromagnetic calorimeter is misidentified as an electron,
and the probability that a minimum ionizing track is misidentified as
a muon. These probabilities are termed fake rates. The fake rate for
each lepton type is calculated using an average of four inclusive jet
samples (triggered with at least one jet with $E_{T} > $ 20, 50, 70, or
100 GeV respectively). We remove sources of real leptons from $Z$ decays using
an invariant mass cut, and from $W$ decays using a Monte Carlo estimate
of the contamination,
and parametrize the fake rates as a function of jet transverse energy
for electrons, or track transverse momentum for muons. The
background is determined by weighting the jets from a data sample of
$(W \rightarrow \ell\nu) + {\rm jets}$ events by the fake rates.
As a result of very low statistics in the data for the calculation of
the fake rates, and large uncertainties in other aspects of the
calculation, we assume a 100\% total uncertainty on our final
$W + {\rm jets}$ background estimates.

A summary of all expected
contributions for each dilepton channel is given in
Table~\ref{table:gstable}, together with observed numbers of
events.  
%There is very good agreement between the
%expected and observed numbers of dilepton events.

%{\bf tables to include: \\
%    -- grand summary table of data and expectations \\

\begin{table}[h]
\caption{\label{table:gstable}
The numbers of SM predicted events, and the numbers observed, in 
$360~{\rm pb^{-1}}$ of data. 
For the $e\mu$ channel, where the only requirement is two
high-$p_T$ leptons, the first three processes are considered signals
for which cross sections are measured. For the $ee$ and $\mu\mu$
channels, where an additional $\metsig$ requirement is made, only the
first two processes are regarded as signals. To calculate the expected
number of events from our signal processes we used the cross section central
values, $\sigma(\tt) = 6.1~\rm{pb}$, $\sigma(\ww) = 12.4~\rm{pb}$, and
$\sigma(\ztt) = 251~\rm{pb}$. Uncertainties on the theoretical cross sections are not
included.}
\begin{ruledtabular}
\begin{tabular}{lcccc}
&   $e\mu$ & $ee$ & $\mu\mu$ & $\ell \ell$  \\ \hline
%\multicolumn{5}{|l|}{``Signal'' processes} \\ 
$\tt$ & $10.0\pm 0.7$ \ & $3.6\pm 0.5$ \ & $3.4 \pm 0.4$ \ & $17.0 \pm 1.6$ \ \\

$\ww$ & $13.8\pm 0.8$ \ & $5.2\pm 0.4$ \ & $4.3 \pm 0.3$ \ & $23.3\pm1.5$ \ \\

$\ztt$ & $57.8\pm4$ \ & $1.1\pm 0.2$ \ & $0.6 \pm 0.1$ \ & $59.5\pm4.3$ \ \\ 
%\multicolumn{5}{|l|}{``Background'' processes} \\ 
$DY \rightarrow ee$ & $0$ \ & $15.4 \pm 3.2$ \ & $0$ \ & $15.4\pm3.2$ \ \\

$DY \rightarrow \mu\mu$  & $9.3\pm 0.8$ \ & $0$ \ & $11.6 \pm 2.4$ \ &
$20.8\pm3.2$ \ \\ 
$WZ$ & $0.70\pm 0.06$ \ & $1.26\pm 0.09$ \ & $1.11 \pm 0.08$ \ & $3.07\pm 0.23$ \
\\ 
$ZZ$ & $0.07\pm 0.01$ \ & $0.47\pm 0.03$ \ & $0.42 \pm 0.03$ \ & $0.96\pm0.07$ \
\\ 
$W\gamma $ & $1.2\pm0.5 $ \ & $1.8\pm 0.7$ \ & $0$ \ & $3.0\pm1.2$ \ \\ 
$W + {\rm jets}$  & $3.0 \pm 3.0$  & $2.1 \pm 2.1$  &  $1.6 \pm 1.6$ & $6.8\pm6.8$\  \\
\hline
%\multicolumn{5}{|l|}{Total expected ``Signal + Background'' event count}\\ 
Total SM   & $96\pm 5$   &  $31\pm 4$  &  $23\pm3$  & $150\pm12$   \\
\hline

Data & $103$  & $24$  & $29$  & $156$  \\
\end{tabular}
\end{ruledtabular}
\end{table}

%---------
% results
%--------

%{\bf discussion of likelihood method here \\}
We extract the $\tt$, $\ww$, and $\ztt$ cross sections simultaneously for the
$e\mu$ channel by maximizing the binned likelihood function:
 
\begin{equation}
L(\mu_{\tt},\mu_{WW},\mu_{\tau\tau})=\prod_i^{N_{bin}}\frac{\mu_i^{n_i}e^{-\mu_i}}{n_i!} 
\times \prod_j  G(x_j, \sigma_j)
\end{equation}
where the index $i$ runs over all $N_{bin}$ bins in the two-dimensional $\met - N_j$
parameter space, and $j$ runs over all variables $x_j$, which are parameters in
the likelihood function constrained by a Gaussian of width $\sigma_j$, the estimated 
uncertainty on $x_j$. These variables consist of the expected number of events
for the background processes (given in Table~\ref{table:gstable}), the acceptances for the
signal processes (given in Table~\ref{table:acctable}), and the integrated luminosity.

The parameters $\mu_{\tt}$, $\mu_{WW}$, and $\mu_{\tau\tau}$ are the expected
total numbers of $\tt$, $\ww$, and $\ztt$ events, respectively. The expected distributions
of these processes in the $\met - N_j$ parameter space determine the probabilities
$p_{\tt,i}$, $p_{WW,i}$, and $p_{\tau\tau,i}$ that a $\tt$, $\ww$, or $\ztt$ event,
respectively, will appear in the $i$-th bin. Therefore, the total expected number
of events in the $i$-th bin of the $\met - N_j$ parameter space is:

\begin{equation}
\mu_i = (p_{\tt,i} \times \mu_{\tt}) + (p_{WW,i} \times \mu_{WW}) + 
 (p_{\tau\tau,i} \times \mu_{\tau\tau})  + n _{other}
\end{equation}
where $n_{other}$ is the expected number of events in the $i$-th bin from all
background processes, and is fixed in the likelihood fit within its Gaussian
constraint. The product over $i$ in the likelihood function is the product
of Poisson probabilities for each bin, in which the expected number of events
is $\mu_i$, and $n_i$ is the corresponding number of events observed in the data. We use 
10 GeV wide bins for $\met<$ 60 GeV, and 20 GeV bins for $60 < \met < 120~{\rm GeV}$. 
Note that in Figs.~\ref{fig:fig1} and~\ref{fig:fig2} we show the $\met - N_j$ shapes 
using a uniform 10 GeV binning in $\met$.

For each of the signal processes, $\mu_{k} = \sigma_{k} \epsilon_{k} {\cal{L}}$, 
where $k = \tt$, $\ww$, or $\ztt$, and $\sigma_{k}$ is the production cross section 
for the process $k$, which is a free parameter in the likelihood fit.
The acceptances, $\epsilon_{k}$, and integrated luminosity of the data sample,
${\cal{L}}$, are fixed within a Gaussian constraint as mentioned above. 
By maximizing the likelihood function we extract the production cross sections for 
$\tt$, $\ww$, and $\ztt$.

We perform a similar likelihood fit to the full $ee + e\mu +
\mu\mu$ data to extract $\tt$ and $\ww$ cross sections, in which we
consider $\ztt$ as a fixed background included in the $n_{other}$
term, since in the $ee$ and $\mu\mu$ channels it has been
significantly reduced by the $\metsig$ requirement. We perform the 
full fit using a likelihood function that is the product of the individual
likelihood functions for each channel. We also make the following assumptions
about correlations in the full fit: within a given channel ($ee$, $e\mu$, or $\mu\mu$)
we assume the signal acceptances are 100\% correlated because they are driven
by the lepton identification efficiencies, and between channels we assume
no correlations in the acceptances. The latter is not actually true because, for example,
the $ee$ and $e\mu$ channels have correlations in the lepton identification uncertainties
due to overlap in lepton types, and the $ee$ and $\mu\mu$
channels share some JES systematic uncertainties. However, by varying
these correlations between 0\% and 100\% we see a negligible effect on
the extracted cross sections, so we 
use 0\% for simplicity.

For both the $e\mu$-only and $ee + e\mu + \mu\mu$ scenarios
we perform two sets of fits. Our main results are obtained by letting
all the signal cross sections float, with the exception
of $\ztt$ in the $ee$ and $\mu\mu$ channels, to
simultaneously extract the signal cross sections from the fit. These results 
are summarized in Table~\ref{table:cstable}, with the systematic
uncertainties included in these results being discussed below. 
As a result of the good separation of our signal processes in the
$\met - N_j$ parameter space, we observe very little correlation 
between these cross section measurements. In the $e\mu$ channel fit
where all three cross sections float, these correlations are about 
$-0.06$ between $\sigma(\tt)$ and $\sigma(\ww)$, $-0.05$ between $\sigma(\tt)$ and 
$\sigma(\ztt)$, and, $-0.19$ between $\sigma(\ww)$ and $\sigma(\ztt)$.
%By letting
%all signal normalizations float we obtain the results most preferred
%globally by the data in our parameter space. 
The $\tt$ production
cross section measurement is relatively insensitive to the top mass
used for the $\tt$ acceptance in generating the $\met - N_j$ template
shape. We generated templates using a top mass of $178\,{\rm GeV}/c^2$~\cite{top178}. 
Using simulated experiments we observe a $1\%$ variation in
$\sigma(t\bar{t})$ if the top mass is varied between $165\,{\rm
GeV}/c^2$ and $178\,{\rm GeV}/c^2$; therefore, we neglect any effects
due to the uncertainty on the top mass.

Fits to the data are also performed by fixing all but one of the signal 
processes to
their SM expected values. We consider the results extracted from these
fits as cross-checks, as they use the added constraint of assuming SM
production for all processes other than the one being measured, in a similar
fashion to the more standard counting experiment results. Significant differences 
between these cross-checks and our main results could be an indication that the
SM assumptions being used are incorrect.
We summarize the results from these fits
in Table~\ref{table:cscc}. When fixed in a particular fit, we use the
following SM theoretical predictions: $\sigma(\tt) = 6.1 \pm 0.9~\rm{pb}$
~\cite{th_ttbar}, $\sigma(\ww) = 12.4 \pm 0.8 ~\rm{pb}$~\cite{th_ww}, and
$\sigma(\ztt) = 251.3 \pm 5.0 ~\rm{pb}$~\cite{th_ztt}.
% ttbar th cross section: 15% error 

The uncertainties on the measured cross sections include a component
coming from the fit (including statistical, acceptance systematic, and integrated 
luminosity) and a second one due to changes in the
$\met - N_j$ distributions caused by the systematic sources mentioned
in Table~\ref{table:accsys}. The fit program used was {\sc minuit}~\cite{minuit},
which minimizes $-\ln (L)$, and determines the fit errors from the 
$-2\ln (L)$ values. 
%The fit procedure and the values returned were 
%checked with simulated experiments assuming SM cross sections and our
%measured acceptances with their systematic uncertainties. 
To evaluate systematic changes in the
shape of $\met - N_j$ distributions, we use simulated experiments with one
of the signal or background distributions from a Monte Carlo simulation with a
particular systematic effect applied. The shape systematic
uncertainties are summarized in Table~\ref{table:shapesyst}.

%   -- table of results \\

\begin{table}[h]
\caption{\label{table:cstable}
Cross section measurements from a global fit of $360 {~\rm pb^{-1}}$ 
of high-$p_T$ dilepton data. The first uncertainty is the error
returned by the likelihood fit, which includes statistical, acceptance
systematic, and luminosity uncertainties. The magnitudes of the
latter two uncertainties are given in Table~\ref{table:accsys} and in 
the text. The second
is the systematic uncertainty in the template shapes.}
\begin{ruledtabular}
\begin{tabular}{lcc}

Process &  $e\mu$ & $ee+\mu\mu+e\mu$ \\
\hline
% $\sigma(\tt)$ ($\ww$, $\ztt$ fixed)  & 
%     $9.3_{-2.6}^{+3.1}(fit)_{-0.2}^{+0.7}(shape)$ pb &  
%     $8.4_{-2.1}^{+2.5}(fit)_{-0.3}^{+0.7}(shape)$  pb\\
$\sigma(\tt)$  &
     $9.3^{+3.1}_{-2.6}\,_{-0.2}^{+0.7}$ pb & 
     $8.5^{+2.6}_{-2.2}\,_{-0.3}^{+0.7}$ pb \\ 
% $\sigma(WW)$  ($\tt$, $\ztt$ fixed)  & 
%     $12.3_{-4.4}^{+5.3}(fit)_{-0.1}^{+0.5}(shape)$ pb &  
%     $16.1^{+5.0}_{-4.3}(fit)_{-0.2}^{+0.8}(shape)$ pb \\
$\sigma(\ww)$   & 
     $11.4^{+5.2}_{-4.3}\,_{-0.1}^{+0.5} $ pb & 
     $16.3^{+5.1}_{-4.4}\,_{-0.2}^{+0.8}$ pb \\ 
% $\sigma(Z\rightarrow\tau\tau)$ ($\tt$, $\ww$ fixed)  & 
%    $292.7^{+48.9}_{-45.1}(fit)^{+5.9}_{-2.9}(shape)$ pb & - \\
     $\sigma(\ztt)$  & 
    $291^{+50}_{-46}\,^{+6}_{-3}$ pb& - \\
\end{tabular}
\end{ruledtabular}
\end{table}

\begin{table}[h]
\caption{\label{table:cscc}
Cross section measurements from a fit to the data with all but
one signal process fixed to its SM value. See text for further details.
The uncertainties have the same meaning as in Table~\ref{table:cstable}.
}
\begin{ruledtabular}
\begin{tabular}{lcc}

Process &  $e\mu$ & $ee+\mu\mu+e\mu$ \\
\hline
 $\sigma(\tt)$ ($\ww$, $\ztt$ fixed)  & 
     $9.3_{-2.6}^{+3.1}\,_{-0.2}^{+0.7}$ pb &  
     $8.4_{-2.1}^{+2.5}\,_{-0.3}^{+0.7}$  pb\\
 $\sigma(\ww)$  ($\tt$, $\ztt$ fixed)  & 
     $12.3_{-4.4}^{+5.3}\,_{-0.1}^{+0.5}$ pb &  
     $16.1^{+5.0}_{-4.3}\,_{-0.2}^{+0.8}$ pb \\
 $\sigma(Z\rightarrow\tau\tau)$ ($\tt$, $\ww$ fixed)  & 
    $293^{+49}_{-45}\,^{+6}_{-3}$ pb & - \\
\end{tabular}
\end{ruledtabular}
\end{table}

\begin{table}[h]
\caption{Summary of the shape systematic uncertainties for the $e\mu$ and full($ee + e\mu + \mu\mu$) 
fit.
\label{table:shapesyst}}
\begin{ruledtabular}
\begin{tabular}{lccccc}

Source  &  $\tt$($e\mu$) & $W^+W^- (e\mu)$  & $\ztt$($e\mu$) & $\tt$
(full) & $W^+W^-$ (full)\\
\hline
JES  & $_{-1}^{+6}\%$  & $_{-1}^{+4}\%$ & $_{-1}^{+2}\%$ & $_{-2}^{+7}\%$ & $_{-1}^{+5}\%$ \\

ISR & $_{-2}^{+4}\%$ & $\pm1\%$ &  $\pm1\%$ &  $_{-2}^{+5}\%$ & $\pm1\%$ \\

FSR  & $\pm1\%$ & --- & --- & $\pm1\%$ & --- \\
\hline
Total & $_{-2}^{+7}\%$ & $_{-1}^{+4}\%$ &  $_{-1}^{+2}\%$ &  $_{-3}^{+8}\%$ & $_{-1}^{+5}\%$ \\
\end{tabular}
\end{ruledtabular}
\end{table}

%-----------
%conclusion
%----------

%{\bf results and conclusion here \\}

In summary, we present a new method to study globally the production
of events with final-states including two high-$p_{T}$ leptons. We
measure simultaneously the production cross sections for $\tt$, $\ww$,
and $\ztt$ and obtain the following results: $\sigma(t\bar{t}) =
8.5^{+2.7}_{-2.2}$ pb, $\sigma(W^+W^-) = 16.3^{+5.2}_{-4.4}$ pb, and 
$\sigma(\ztt) = 291^{+50}_{-46}$ pb. They are in good agreement 
with SM theoretical predictions. 
In addition to the potential this analysis technique has for precision 
cross section measurements in the dilepton channel with more data, it could 
also be promising for model independent searches for new physics.
%
%This analysis represents the first step
%towards looking for new physics in the high-$p_T$ dilepton sample
%using a global approach that maximizes sensitivity and minimizes model
%dependence.
%

%------------------------------------------------------------------

%-----------
%Acknowledgments
%-----------
We thank the Fermilab staff and the technical staffs of the participating 
institutions for their vital contributions. This work was supported by 
the U.S. Department of Energy and National Science Foundation; the 
Italian Istituto Nazionale di Fisica Nucleare; the Ministry of Education, 
Culture, Sports, Science and Technology of Japan; the Natural Sciences 
and Engineering Research Council of Canada; the National Science Council 
of the Republic of China; the Swiss National Science Foundation; the A.P. 
Sloan Foundation; the Bundesministerium f\"ur Bildung und Forschung, 
Germany; the Korean Science and Engineering Foundation and the Korean 
Research Foundation; the Particle Physics and Astronomy Research Council 
and the Royal Society, UK; the Institut National de Physique Nucleaire et 
Physique des Particules/CNRS; the Russian Foundation for Basic Research; 
the Comisi\'on Interministerial de Ciencia y Tecnolog\'{\i}a, Spain;
the European Community's Human Potential Programme under contract 
HPRN-CT-2002-00292; and the Academy of Finland.

%We thank the Fermilab staff and the technical staffs of the participating institutions 
%for their vital contributions. This work was supported by the U.S. Department of Energy 
%and National Science Foundation; the Italian Istituto Nazionale di Fisica Nucleare; the 
%Ministry of Education, Culture, Sports, Science and Technology of Japan; the Natural Sciences 
%and Engineering Research Council of Canada; the National Science Council of the Republic of China; 
%the Swiss National Science Foundation; the A.P. Sloan Foundation; the Bundesministerium f\"ur Bildung 
%und Forschung, Germany; the Korean Science and Engineering Foundation and the Korean Research 
%Foundation; the Particle Physics and Astronomy Research Council and the Royal Society, UK; the 
%Russian Foundation for Basic Research; the Comisi\'on Interministerial de Ciencia y 
%Tecnolog\'{\i}a, Spain; in part by the European Community's Human Potential Programme under 
%contract HPRN-CT-2002-00292; and the Academy of Finland. 

%-----------------------
%  Bibliography
%-----------------------


\begin{thebibliography}{99}

\bibitem{dilcross} F. Abe {\it et al.} (CDF Collaboration), Phys. Rev. Lett. {\bf 80}, 2779 (1998).

\bibitem{barney}R. M. Barnett and L. J. Hall, Phys. Rev. Lett. {\bf 77}, 3506 (1996). 

%\bibitem{dilkin}  D. Acosta {\it et al.} (CDF Collaboration), Phys. Rev. Lett. {\bf 95}, 022001 (2005).

%\bibitem{ditau} M. Carena  {\it et al.}, Phys. Rev. D {\bf 70}, 093009 (2004).
% from ditau paper hep-ex/0506034

\bibitem{topcolor} C. T. Hill, Phys. Lett. B {\bf 266}, 419 (1991).
%http://www-cdf.fnal.gov/physics/new/top/confNotes/cdf8087_mtt680.ps

\bibitem{4gen}E. Arik {\it et al.}, Phys. Rev. D {\bf 66}, 033003 (2002).

\bibitem{hww}A. Abulencia {\em et al.} (CDF Collaboration), Phys. Rev. Lett. {\bf 97}, 081802 (2006).

\bibitem{peetee} In the CDF coordinate system, $\theta$ and $\phi$ are the 
polar and azimuthal angles, respectively, with respect to the proton
beam direction ($z$ axis). The pseudorapidity $\eta$ is defined as
$-\ln \tan(\theta/2)$.  The transverse momentum of a particle is $p_T
= p \sin\theta$.  The analogous quantity using calorimeter energies,
defined as $\Et = E \sin\theta$, is called transverse energy.  
%
%\bibitem{jetcorr} G. C. Blazer and B. L. Flaugher, Ann. Rev. of Nucl. and Part. Sci. Vol. {\bf 49} (1999).

\bibitem{met} The missing transverse energy, $\vec{\met}$, is defined as $-\sum \Et^i
\,\hat{n}_i$, where $\hat{n}_i$ is the unit vector in the transverse
plane pointing from the interaction point to the energy deposition in
calorimeter cell $i$. The $\vec{\met}$ measurement is corrected for
muons in the event with $p_{T}> 20~{\rm GeV}/c$, and for jet
corrections. The magnitude of $\vec{\met}$ is denoted as $\met$.

\bibitem{ttdilxsec} D. Acosta {\it et al.} (CDF Collaboration), Phys. Rev. Lett. {\bf 93}, 142001 (2004).

\bibitem{WWxsec} D. Acosta {\it et al.} (CDF Collaboration), Phys. Rev. Lett. {\bf 94}, 211801 (2005).

\bibitem{zttxsec} A. Safonov,  Nucl. Phys. Proc. Suppl. {\bf 144}, 323-332 (2005).

\bibitem{cdfdet}D. Acosta {\it et al.} (CDF Collaboration), Phys. Rev. D {\bf 71}, 032001 (2005).

\bibitem{jetcorr} G. C. Blazey and B. L. Flaugher, Ann. Rev. of Nucl. and Part. Sci. Vol. {\bf 49} (1999).

\bibitem{jes} A. Bhatti {\it et al.}, Nucl. Instrum. Methods A {\bf 566}, 375 (2006).

\bibitem{trigele} A. Abulencia  {\it et al.}(CDF Collaboration), hep-ex/0508029, submitted to Phys. Rev. D.

%\bibitem{trigplug} D. Acosta {\it et al.} (CDF Collaboration), Phys. Rev. D {\bf 71} (Rapid Communications), 051104 (2005).

%\bibitem{wz_prd} W/Z PRD -- get reference.

\bibitem{caliso} The calorimeter isolation $I_{cal}$ is defined as the extra energy deposited in the 
 calorimeter cone of radius 
$\Delta R=\sqrt{{\Delta \phi}^{2}+ {\Delta \eta}^{2}}=$ 0.4 around the lepton. The $I_{cal}$ is required 
to be less than 
10 $\%$ of the lepton $E_{T}$.

\bibitem{trkiso} The track isolation $I_{trk}$ is defined as the sum of the $p_T$ of all the tracks in a 
cone of radius 
$\Delta R=\sqrt{{\Delta \phi}^{2}+ {\Delta \eta}^{2}}=$ 0.4 around the lepton 
candidate track, but excluding it. The $I_{trk}$ is required to be less than 10$\%$ of the lepton track $p_{T}$. 

\bibitem{pythia}T. Sj\"{o}strand {\it et al.}, Comput. Phys. Commun. {\bf 135}, 238 (2001). We use 
{\sc pythia} V6.2.

\bibitem{geant} R. Brun and F. Carminati, CERN Program Library Long Writeup W5013, 1993
(unpublished).

%\bibitem{jes} A. Bhatti {\it et al.}, Nucl. Instrum. Methods A {\bf 566}, 375 (2006).

\bibitem{pdferror} We compared the acceptances obtained using MRST 72 and 75 PDF sets~\cite{mrst},
corresponding to $\alpha_{s} =$ 0.1175 and 0.1125 respectively.  Also we
include the effect of the eigenvector variations around the fit
minima, using CTEQ6M PDF sets~\cite{cteq}, and used the maximum variation observed 
in the acceptance as our systematic uncertainty.

\bibitem{mrst} A. D. Martin, R. G. Roberts, W. J. Stirling, and R. S. Thorne, hep-ph/0307262.

\bibitem{cteq} H. L. Lai {\it et al.}, Phys. Rev. D {\bf 51}, (1995).

\bibitem{lum} S. Klimenko, J. Konigsberg, and T.M. Liss, FERMILAB-FN-0741 (2003).
%D. Acosta {\it et al.}, Nucl. Instrum. Methods A {\bf 461}, 540 (2001). 

\bibitem{wwxsec}J. M. Campbell and R. K. Ellis, Phys. Rev. D {\bf 60}, 113006 (1999).

\bibitem{baur1} U. Baur and E. L. Berger, Phys. Rev. D {\bf 41}, 1476 (1990); U. Baur and E. L. Berger, 
Phys. Rev. D {\bf 47}, 4889 (1993).

\bibitem{top178} P. Azzi {\it et al.} (for the CDF and D\O\ collaborations), hep-ex/0404010; 
superseded by hep-ex/0608032.

\bibitem{th_ttbar} R. Bonciani, S. Catani, M. L. Mangano, and P. Nason, Nucl. Phys. B {\bf 529}, 424 (1998), 
update in hep-ph/0303085.

\bibitem{th_ww} J. M. Campbell and R. K. Ellis, Phys. Rev. D {\bf 60} (1999) 113006; 
next-to-leading-order calculation using MCFM version 3.4.5.

\bibitem{th_ztt} A. D. Martin, R. G. Roberts, W. J. Stirling, and R. S. Thorne, 
  Eur. Phys. J. C {\bf 35}, 325-348 (2004).

\bibitem{minuit} F. James, CERN Program Library Long Writeup D506, {\sc minuit} reference manual version 94.1,
1998 (unpublished).

\end{thebibliography}
\end{document}